\documentstyle[12pt]{article}

\begin{document}
\begin{center}
{\Large \bf Scenario of baryogenesis }
\bigskip

{\large D.L.~Khokhlov}
\smallskip

{\it Sumy State University, R.-Korsakov St. 2, \\
Sumy 244007, Ukraine\\
E-mail: others@monolog.sumy.ua}
\end{center}

\begin{abstract}
Scenario of baryogenesis is considered in which
primordial plasma starting from the Planck scale
consists of primordial particles being
the precursors of electrons and clusters of particles being
the precursors of protons.
Equilibrium between the precursors of protons and the precursors of
electrons is defined by the proton-electron mass difference.
At the temperature equal to the mass of electron,
primordial particles transit into protons, electrons, photons.
\end{abstract}

Standard scenario of baryogenesis occurs as follows~\cite{Kolb}.
In the thermodynamically equilibrium primordial plasma,
all the particles with $m\leq T$ exist in equal abundances
per spin degree of freedom. This means that, at $T>m_{p}$,
there exist approximately equal abundances of protons $p$ and
antiprotons $\bar p$. Excess of baryonic charge arises in the
processes of X, Y-bosons decays under the following conditions:
non-conservation of the baryonic charge and CP-violation.
At $T<m_{p}$, $p\bar p$ annihilate, with the excess of $p$
survives.

Let us consider another scenario of baryogenesis in which
primordial plasma starting from the Planck scale $T=m_{Pl}$
consists of primordial fermions.
At $T=m_{e}$, primordial fermions transit into protons, electrons,
photons.

Let us assume that primordial plasma consists of primordial
fermions which have only spin quantum number. That is
primordial plasma consists of primordial fermions with the spin
up $\uparrow$ and down $\downarrow$.
Then the sequence of transformations
$\uparrow\downarrow\uparrow\downarrow\uparrow$
transits the particle into itself.
Let us identify simple particle $\uparrow$ with the precursor
of electron and the cluster of particles
$\uparrow\downarrow\uparrow\downarrow\uparrow$
with the precursor of proton.
The probability of finding such a cluster is given by
\begin{equation}
w=\left(\frac{1}{2}\right)^{5}.
\label{eq:w}
\end{equation}
Equilibrium between the precursors of protons and the precursors of
electrons is defined by the proton-electron mass difference.
The proton-electron ratio is given by
\begin{equation}
\frac{N_{p}}{N_{e}}=\left(\frac{1}{2}\right)^{5}
\left(\frac{m_{e}}{m_{p}}\right)^{2}
\label{eq:NpNe}
\end{equation}

Let us assume that, at $T=m_{e}$, precursors of electrons
annihilate into photons, and precursors of protons transit into
protons, with the same number of precursors of electrons transit
into electrons. Extraction of protons and electrons leads to
the appearance of the electric charge.
Baryon-photon ratio at $T=m_{e}$ is given by
\begin{equation}
\frac{N_{b}}{N_{\gamma}}=\frac{1}{2}\times\frac{3}{4}\times\frac
{N_{p}}{N_{e}}
\label{eq:NbNg}
\end{equation}
where fraction $1/2$ takes into account survived electrons,
and fraction $3/4$ takes into account relation between
fermions and bosons. Calculations yield the value
$N_{b}/N_{\gamma}=3.5\times 10^{-9}$.

The observed value of $N_{b}/N_{\gamma}$ lies in the range
$2-15\times 10^{-10}$~\cite{ISSI}.
Let us assume that the most fraction of baryonic matter
decays into non-baryonic matter during the evolution of the universe.
Estimate baryon number density at $T=m_{e}$ from the modern total
mass density of the universe. According to the model of the
universe with the linear evolution law~\cite{Kh},
mass density of the universe is given by
\begin{equation}
\rho={3\over{4\pi G t^2}}.
\label{eq:rho}
\end{equation}
Modern age of the universe is given by
\begin{equation}
t_{0}=t_{Pl}\alpha\left(\frac
{T_{Pl}}{T_{0}}\right)^2.
\label{eq:age}
\end{equation}
From this the modern age of the universe is equal to $t_{0}=
1.06\times 10^{18} \ {\rm s}$,
and the modern mass density of the universe is equal to
$\rho_{0}=3.19\times 10^{-30}\ {\rm g\ cm^{-3}}$.
Then the baryon number density at $T=m_{e}$ is
$n_{b}=\rho_{0}/m_{p}=1.9\times 10^{-6}\ {\rm cm^{-3}}$.
While adopting the observed photon number density as
$n_{\gamma}=550 \ {\rm cm^{-3}}$~\cite{Dolg},
the baryon-photon ratio at $T=m_{e}$ is equal to
$N_{b}/N_{\gamma}=3.5\times 10^{-9}$.

In the standard theory of primordial nucleosynthesis~\cite{Walk},
neutron-proton ratio freezes out at
$T\approx 1 \ {\rm MeV}$.
If baryons arise at $T=m_{e}$, neutron-proton ratio freezes out at
$T\leq m_{e}\approx 0.5 \ {\rm MeV}$ that leads to the decrease of
the freezed out neutron-proton ratio.


\begin{thebibliography}{99}
\bibitem{Kolb}
E.W.~Kolb, M.S.~Turner, Ann. Rev. Nucl. Part. Sci. {\bf V.~33}
(1983) 645; M.S.~Turner in {\em Architecture of Fundamental
Interactions at Short Distances}, eds. P.~Ramond and R.~Stora
(Elsevier Sci. Publ., Copenhagen, 1987)

\bibitem{ISSI}
{\em ISSI Workshop on Primordial Nuclei and Their Evolution}
(Bern, 1997), ed. N.~Prantzos, M.~Tosi, and R.~von~Steiger
(Kluwer, Dordrecht)

\bibitem{Kh}
D.L. Khokhlov, astro-ph/9811311

\bibitem{Dolg}
A.D.~Dolgov, Ya.B.~Zeldovich, M.V.~Sazhin, {\em Cosmology of the
early universe} (Moscow Univ. Press, Moscow, 1988, in Russian).

\bibitem{Walk}
T.P.~Walker, G.~Steigman, D.N.~Schramm, K.A.~Olive, and H-S.~Kang,
Ap. J. {\bf 376} (1991) 51

\end{thebibliography}
\end{document}